\documentclass[twocolumn]{aastex61}
\usepackage{amsmath}
\graphicspath{{Img/},{./}}

\begin{document}
\title{Joining X-ray to lensing: an accurate combined analysis of MACS J0416.1$-$2403}

\author{M.~Bonamigo}
\affiliation{Dark Cosmology Centre, Niels Bohr Institute, University of Copenhagen, Juliane Maries Vej 30, DK-2100 Copenhagen, Denmark} 

\author{C.~Grillo}
\affiliation{Dipartimento di Fisica, Universit\`a degli Studi di Milano, via Celoria 16, I-20133 Milano, Italy} 
\affiliation{Dark Cosmology Centre, Niels Bohr Institute,
University of Copenhagen, Juliane Maries Vej 30, DK-2100 Copenhagen, Denmark} 

\author{S.~Ettori}
\affiliation{INAF, Osservatorio Astronomico di Bologna, Via Piero Gobetti, 93/3, 40129 Bologna, Italy}
\affiliation{INFN, Sezione di Bologna, viale Berti Pichat 6/2, 40127 Bologna, Italy}

\author{G.~B.~Caminha}
\affiliation{Dipartimento di Fisica e Scienze della Terra, Universit\`a degli Studi di Ferrara, Via Saragat 1, I-44122 Ferrara, Italy} 

\author{P.~Rosati}
\affiliation{Dipartimento di Fisica e Scienze della Terra, Universit\`a degli Studi di Ferrara, Via Saragat 1, I-44122 Ferrara, Italy} 

\author{A.~Mercurio}
\affiliation{INAF - Osservatorio Astronomico di Capodimonte, Via Moiariello 16, I-80131 Napoli, Italy} 

\author{M.~Annunziatella}
\affiliation{INAF - Osservatorio Astronomico di Trieste,  Via G.B. Tiepolo, 11 I-34143 Trieste, Italy} 

\author{I.~Balestra}
\affiliation{University Observatory Munich, Scheinerstrasse 1, D-81679 Munich, Germany}

\author{M.~Lombardi}
\affiliation{Dipartimento di Fisica, Universit\`a degli Studi di Milano, via Celoria 16, I-20133 Milano, Italy}

\email{bonamigo@dark-cosmology.dk}


\begin{abstract}
We present a novel approach for a combined analysis of X-ray and gravitational
lensing data and apply this technique to the merging galaxy cluster MACS J0416.1$-$2403.
The method exploits the information on the intracluster gas distribution that
comes from a fit of the X-ray surface brightness and then includes the hot gas
as a fixed mass component in the strong-lensing analysis.
With our new technique, we can separate the collisional from the collision-less
diffuse mass components, thus obtaining a more accurate reconstruction of the
dark matter distribution in the core of a cluster.
We introduce an analytical description of the X-ray emission coming from a set
of dual pseudo-isothermal elliptical mass distributions,
which can be directly used in most lensing softwares.
By combining \emph{Chandra} observations with Hubble Frontier Fields imaging and
Multi Unit Spectroscopic Explorer spectroscopy in MACS J0416.1$-$2403, we measure a projected gas-to-total
mass fraction of approximately $10\%$ at $350$ kpc from the cluster center.
Compared to the results of a more traditional cluster mass model (diffuse halos
plus member galaxies), we find a significant difference in the cumulative
projected mass profile of the dark matter component and that the dark matter over
total mass fraction is almost constant, out to more than $350$ kpc.
In the coming era of large surveys, these results show the need of multiprobe
analyses for detailed dark matter studies in galaxy clusters.
\end{abstract}


\section{Introduction} \label{sec:intro}
Galaxy clusters are one of the most powerful and promising tools available for the
study of the different mass components of the Universe \citep{Voit2005,Jullo2010,Kneib2011,Laureijs2011,Postman2012}.
They are the largest gravitationally bound objects in the sky, and as such, they
represent a young population that formed only recently, in accordance with the
hierarchical assembly predicted by the concordance cosmological model
\citep[e.g.,][]{Tormen1997,Klypin1999,Moore1999,Springel2001}.
The late formation time of clusters makes their mass and number density
distributions sensitive to the presence of dark energy, which only recently
has been dominating the dynamical evolution of the Universe.
Galaxy clusters are also the strongest gravitational
lenses, with dozens of families of observed multiple images of background
sources \citep{Broadhurst2005,Halkola2006}.
In addition, galaxy clusters host very hot gas halos arising from the infall of surrounding
material into their deep potential wells \citep{Sarazin1988,Ettori2013}.
For these reasons, galaxy clusters have been targeted in many observational
campaigns, using various techniques and facilities, to study their different
mass components.
For example, the Cluster Lensing And Supernova survey with Hubble
\citep[CLASH;][]{Postman2012}, the Hubble Frontier Fields \citep[HFF;][]{Lotz2017}
and the Reionization Lensing Cluster Survey (RELICS\footnote{\url{https://relics.stsci.edu/}})
have obtained multiband \emph{Hubble Space Telescope (HST)} images, supplemented with ground-based
telescope photometric and spectroscopic data, to map the cluster total mass distribution via strong and weak
gravitational lensing.
Targeted observations using X-ray telescopes, like \emph{Chandra} and \emph{XMM Newton},
and submillimeter (for the Sunyaev–Zel'dovich effect) and radio antennas have
characterized the hot cluster gas component
\citep[e.g.,][]{Donahue2014,Ogrean2016,Rumsey2016,vanWeeren2017}.
Thousands of member galaxies and multiply lensed images have been
spectroscopically confirmed using the Visual Multi-Object Spectrograph
\citep{LeFevre2003}, for the CLASH-VLT program \citep{Rosati2014}, and
more recently the Multi Unit Spectroscopic Explorer \citep[MUSE;][]{Bacon2012} 
instruments at the VLT \citep[e.g.,][]{Biviano2013,Karman2017,Monna2017}.
All these campaigns have helped to push further our understanding of the mass
composition of galaxy clusters and the population of high-$z$ sources.
In particular, the HFF program has dedicated 140 \emph{HST} orbits to each of
the six massive clusters in the sample with the aim of enabling the study of the
population of the highest-redshift galaxies, the first to undergo star
formation, thanks to the magnification effect of the cluster lenses
\citep[e.g.,][]{Balestra2013,Coe2015,Oesch2015,Vanzella2017}.
Moreover, the HFF data represent an unprecedented opportunity to improve the
lensing modeling of the clusters and thus to study in more detail the dark
matter halos in which they live.

In this work, we focus on the galaxy cluster MACS J0416.1$-$2403 (hereafter MACS
0416), present in both the CLASH and HFF samples and first discovered in the Massive
Cluster Survey (MACS) by \citet{Mann2012}.
MACS~0416 is located at $z_l = 0.396$ and has an $M_{200}$ mass of approximately $9 \times 10^{14} M_\odot$
and an X-ray luminosity $L_X \approx 10^{45}$ erg s$^{-1}$ \citep{Balestra2016}.
The cluster hosts two brightest cluster galaxies (BCGs), G1 and G2, located,
respectively, in its northeast (R.A. $=$ 04:16:09.154 and decl. $= -$24:04:02.90)
and southwest (R.A. $=$ 04:16:07.671 and decl. $= -$24:04:38.75) regions.
MACS~0416 is clearly undergoing a merging event, as is shown by its X-ray
morphology and by the large separation ($\sim 200$ kpc) in projection of the two
BCGs \citep{Mann2012}.

Given its inclusion in the HFF sample and its high efficiency in magnifying
background sources, MACS~0416 has been the target of many recent studies.
After the first strong-lensing analysis by \citet{Zitrin2013},
which identified 70 multiple images, \citet{Jauzac2014,Jauzac2015} combined
strong- and weak-lensing data to model a total of $194$ multiple images (almost
all of them without spectroscopic confirmation).
Additional works \citep{Johnson2014,Richard2014,Diego2015} focused on this cluster
and provided maps of its total mass and magnification factors.
Exploiting the spectra obtained within the CLASH-VLT program 
\citep[presented in][]{Balestra2016}, \citet[hereafter Gr15]{Grillo2015}
accurately modeled the positions of $30$ multiple images, all from spectroscopically confirmed sources.
More recently, \citet[][hereafter Ca17]{Caminha2017} used HFF and MUSE data to improve the lensing analysis by Gr15,
extending the number of secure spectroscopic multiple images to $102$ and making MACS~0416 the cluster
with the largest number of spectroscopically confirmed multiple images known to date.

Complementing the gravitational lensing analysis of Gr15, \citet{Balestra2016}
used X-ray data and the dynamics of approximately $800$ member galaxies to
independently measure the total mass of the cluster, finding a good agreement
between the different mass diagnostics.
Similarly, \citet{Jauzac2015} used X-ray data to model the hot gas
distribution in order to clarify the merging history of MACS~0416.
Both groups treated X-ray and strong-lensing data separately.
Ideally, one would want to simultaneously fit the gravitational lensing and X-ray
data, combining the individual likelihoods into a single value to maximize.
This has been done by various authors, most recently by \citet{Morandi2012},
\citet{Umetsu2015}, \citet{Siegel2016}, and \citet{Sereno2017}; for a more detailed review on the
subject see \citet{Limousin2013}.
However, the downside of the current implementations of this approach is that the
lensing analysis is done separately and the observable that enters in the
combined fit is the fixed, reconstructed total surface mass density.
An alternative, and complementary, strategy is to measure the hot gas mass density from the X-ray surface
brightness and then include it in a proper strong-lensing analysis that uses
the positions of observed multiple images as constraints.
This is the method we choose in the current paper to improve the strong
gravitational lensing analysis of Ca17 by adding multiwavelength
information from the X-ray emission of the hot intracluster gas.
This approach has some advantages over a more traditional analysis,
where the hot gas is subtracted a posteriori from the diffuse halo component.
In particular, the inferred dark matter mass density distribution could differ because of the
added constraints from the X-ray data. Moreover, a traditional analysis cannot
measure the parameter values of the diffuse dark matter halo without the bias
introduced by the hot intracluster gas.
A similar technique has been attempted by \citet{Paraficz2016} in the Bullet
cluster (1E 0657$-$56), where the large offset between the X-ray emission and
the total mass distribution required the separate treatment of the hot gas
component in the gravitational lensing analysis.
The wealth of data available for MACS~0416 allows us here to adopt a
much more accurate description for the hot gas and the other components of the cluster,
for example, by modeling the spatial distribution of the X-ray emission beyond
the approximation of a single mass density profile.

There are three main reasons to combine information from different mass diagnostics:
first, any systematic effect (or absence of it) should become evident in the
disagreement of the probes considered \citep{Balestra2016}; secondly, as the
various datasets depend differently on each component, some degeneracies can be
broken \citep[for instance, projection effects;][]{Limousin2013}; lastly, a
multiwavelength analysis can help separate the constituents of a cluster,
allowing for a more detailed study of the individual components.
This is key to testing the collisionless nature of dark matter in
merging systems and the inner structure of dark matter halos predicted by the
standard cold dark matter (CDM) model.
For instance, in the presence of self-interacting dark matter (SIDM), the three main
components of a cluster (dark matter, hot intracluster gas, and member galaxies)
should exhibit a precise displacement after the first
passage in a merging event \citep{Markevitch2004,Harvey2015}.
The center of the galaxies' distribution, in each subcluster, should be located
farther away from the overall barycenter, as they represent a fully
noncollisional component that moves almost unperturbed through the cluster.
The opposite behavior is characteristic of the hot intracluster gas, which, being
a collisional fluid, is compressed and lags behind during the core passage.
The center of an SIDM distribution should be found
somewhere in the middle between the other two components.
In principle, it is possible to constrain the cross section of SIDM from the offset
between the center positions of the dark matter and galaxies' distributions
\citep{Markevitch2004,Harvey2015}.
We notice that in previous studies about MACS 0416 \citep[e.g.,][]{Ogrean2015}
only a very small displacement has been observed between the centers of the
X-ray and optical luminosity peaks.
This might be ascribed to the complex cluster merging geometry, which renders
also an estimate of the DM cross section less straightforward.
To this last particular aim, the technique presented here should be more
effective in galaxy clusters with more favorable geometrical configurations.

The paper is organized as follows.
In Section \ref{sec:xray_analysis}, we introduce the X-ray observations and the
modeling technique used to estimate the hot intracluster gas.
In Section \ref{sec:lensing_analysis}, we briefly present the strong-lensing data
and the adopted cluster mass models used in our analysis.
Section \ref{sec:results} contains the results of the strong-lensing study,
where the hot gas is treated as a separate and fixed mass component.
Finally, in Section \ref{sec:conclusions}, we summarize our conclusions.

Throughout the paper, we adopt a flat $\Lambda$CDM cosmology with Hubble
constant $H_0 = 70$ km s$^{-1}$ Mpc$^{-1}$ and total matter density
$\Omega_m = 0.3$.
At the redshift of the lens of $z_{l} = 0.396$, $1 \arcsec$ corresponds to
$5.34$ kpc in the assumed cosmology.
All magnitudes are given in the AB system.

\section{X-Ray surface brightness analysis}
\label{sec:xray_analysis}
In this section, we present our new technique to model the hot gas mass
distribution of a cluster, and apply this method to MACS~0416.

\subsection{X-Ray Surface Brightness from Dual Pseudo-isothermal Elliptical Mass Density Profiles}
The X-ray surface brightness $S_X(x,y)$ of an object at redshift $z_l$ is given by
\begin{equation}
S_X(x,y) = \frac{\Lambda(T,Z)}{4 \pi (1+z_l)^4} \int_{-\infty}^{+\infty}\!\!\! n_e(x,y,z) n_p(x,y,z)\, \text{d}z,
\label{eq:surface_brightness}
\end{equation} 
where $n_e(x,y,z)$ is the electron density, $n_p(x,y,z)$ is the proton density,
$Z$ is the metallicity of the gas, and $\Lambda(T,Z)$ is its cooling function
\citep{Boehringer1989,Sutherland1993}.

Traditionally, two methods are used to derive the hot gas density from the X-ray surface
brightness \citep[see][and references therein]{Ettori2013}: (1) by considering the geometry of the system, it is
possible to deproject the surface brightness and obtain the gas
density; (2) modeling the gas density and then projecting it allows one
to fit the observed X-ray photon counts and thus infer the parameter values of the assumed
gas distribution.
In this work, we adopt the second approach (because the multipeak mass 
distribution of MACS~0416 makes a simple geometrical deprojection less suitable).
Moreover, we explicitly seek an analytical description of the hot gas mass
density distribution that can be easily included in strong-lensing models.
To this aim, we adopt a dual pseudo-isothermal elliptical (dPIE) mass 
distribution \citep{Eliasdottir2007,Suyu2010}, largely used in
strong-lensing analyses, instead of a $\beta$-model \citep{Cavaliere1976,Sarazin1977},
more common in X-ray studies, to describe the hot gas.
The three-dimensional mass density of a dPIE distribution, with vanishing ellipticity,
can be expressed as \citep{Limousin2005}
\begin{equation}
\rho(x,y,z) = \frac{\rho_{0}}{
    \left( 1 + \frac{ x^2 + y^2 + z^2 }{ R_{C}^2 } \right)
    \left( 1 + \frac{ x^2 + y^2 + z^2 }{ R_{T}^2 } \right)
},
\label{eq:dpie_3Ddens}
\end{equation}
where $\rho_0$ is the central density and $R_C$ and $R_T$ are the core and
truncation radii, respectively.
A dPIE distribution can be seen as a special case of the density profile introduced
by \citet{Vikhlinin2006} as a generalization of the $\beta$-model.
The value of $\rho_0$ is related to that of the normalization of the surface
mass density, i.e.\ the ``central velocity dispersion'' $\sigma_0$ in Equation (\ref{eq:dpie_kappa}), via
\begin{equation}
    \rho_0 = \frac{\sigma_0^2}{2 \pi G} \left( \frac{R_C + R_T}{R_C^2R_T} \right).
\end{equation}
We notice that here the term ``velocity dispersion'' does not have a dynamical
meaning and has to be considered only as an effective parameter.
Hereafter, we will use the following substitutions:
\begin{equation}
\left\{
\begin{array}{lr}
    R_A^2 = R_{C}^2 + (x-x_0)^2 + (y-y_0)^2\\
    R_B^2 = R_{T}^2 + (x-x_0)^2 + (y-y_0)^2\\
\end{array}
\right. ,
\label{eq:dpie_radii}
\end{equation}
thus shifting the center of the profile at the position $(x_0,y_0)$ on the plane of the
sky.

Neglecting constant factors, such as the conversion from electron and proton
densities to gas density, and using Equation (\ref{eq:dpie_3Ddens}), we obtain the
following analytical solution for the surface brightness shown in Equation
(\ref{eq:surface_brightness}):
\begin{equation}
\begin{split}
S_X & \propto \int_{-\infty}^{+\infty}\!\!\! \rho^2 \text{d}z
= \int_{-\infty}^{+\infty}\!\!\! \frac{ \rho_{0}^2 R_{C}^4 R_{T}^4  \text{d}z }{
         \left( R_{A}^2 + z^2 \right)^2
         \left( R_{B}^2 + z^2 \right)^2
    } = \\
& = \frac{ \pi \rho_{0}^2 R_{C}^4 R_{T}^4 \left(  R_{A}^2 + 3 R_{A} R_{B} + R_{B}^2 \right)}
{ 2 R_{A}^3 R_{B}^3  \left(R_{A} + R_{B}\right)^3 }.
\end{split}
\label{eq:I_i}
\end{equation}

It is possible to generalize the problem to the situation where multiple
components are present, but we need to assume that they all lie on the
same plane along the line of sight to obtain an analytical result.
The inclusion of a possible difference in $z$ between separate gas components is beyond the scope of this analysis
and not relevant for the strong-lensing analysis of MACS~0416.
Within this single-plane assumption, the surface brightness of $N$ dPIE components is proportional
to
\begin{equation}
\int_{-\infty}^{+\infty}\!\!\! \rho^2 \text{d}z = 
\sum_{i = 1}^{N} I_i(x,y) + 2 \sum_{i \neq j}^{N} I_{i,j}(x,y),
\label{eq:dNPIE}
\end{equation}
where $I_i(x,y)$ is the solution to the one-component problem (in Equation \ref{eq:I_i})
and the second term $I_{i,j}(x,y)$ is
\begin{equation}
\begin{split}
&I_{i,j}(x,y) = \int_{-\infty}^{+\infty}\!\!\! \rho_i \rho_j \, \text{d}z = \\
& = \int_{-\infty}^{+\infty}\!\!\! \frac{ \left( \rho_{0i} R_{Ci}^2 R_{Ti}^2 \right)
\left( \rho_{0j} R_{Cj}^2 R_{Tj}^2 \right) \text{d}z }{
         ( R_{Ai}^2 + z^2 )
         ( R_{Bi}^2 + z^2 )
         ( R_{Aj}^2 + z^2 )
         ( R_{Bj}^2 + z^2 ) } = \\
& = \pi \left( \rho_{0i} R_{Ci}^2 R_{Ti}^2 \right)
\left( \rho_{0j} R_{Cj}^2 R_{Tj}^2 \right) \frac{\alpha_{i,j}}{\beta_{i,j}}\; ,
\end{split}
\end{equation}
where
\begin{equation}
\small
\begin{split}
&\alpha_{i,j} = (R_{Bi}+R_{Aj}) (R_{Bi}+R_{Bj}) (R_{Aj}+R_{Bj}) + \\
& +  R_{Ai} (R_{Bi}+R_{Aj}+R_{Bj})^2  + R_{Ai}^2 (R_{Bi}+R_{Aj}+R_{Bj})
\end{split}
\end{equation}
and
\begin{equation}
\small
\begin{split}
&\beta_{i,j} = R_{Ai} R_{Bi} R_{Aj} R_{Bj} (R_{Ai}+R_{Bi}) (R_{Ai}+R_{Aj}) \times \\
& \times (R_{Ai}+R_{Bj}) (R_{Bi}+R_{Aj}) (R_{Bi}+R_{Bj}) (R_{Aj}+R_{Bj}).
\end{split}
\end{equation}
In appendix \ref{sec:d2PIE}, we show the solution for the particular case of two
components and we also discuss a generalization beyond the spherically
symmetric approximation.

From an X-ray only fit, it is not possible to break the degeneracy 
between surface brightness normalization and elongation along the line of
sight; therefore, the surface mass density measured under the spherical approximation
is biased by a factor that depends on the real geometry of the system.
In the most unfavorable scenario of a prolate
ellipsoid, with axis ratio $s$ aligned with the line of sight, this factor is $1/\sqrt{s}$ \citep{DeFilippis2005}.
Without further information from other observables, though, this bias cannot be
quantified in MACS~0416, especially given the extremely
complex nature of this cluster.

\subsection{X-Ray Surface Brightness Fit of MACS~J0416.1$-$2403}
First, we model the X-ray surface brightness of MACS~0416.
We combine multiple \emph{Chandra} observations \citep[obsID: 16236, 16237, 16304, 16523,
17313; see][]{Ogrean2015}, for a total of 293 ks of exposure time, and reduce
them using \emph{CIAO 4.7} and \emph{CALDB 4.6.9}.
The resulting surface brightness map, limited in the energy range from $0.7$ to
$2$ keV and corrected for exposure, is then binned to 8 times the pixel
resolution of \emph{Chandra}, obtaining a final pixel size of $3\arcsec.94$.
This pixel size is much larger than \emph{Chandra}'s on-axis point-spread function;
therefore, we do not consider this effect in our analysis.

From the modeling presented in the previous section, we can obtain the projected squared
gas density (see Equation \ref{eq:dNPIE}).
To convert this into an X-ray surface brightness, we use Equation~(\ref{eq:surface_brightness}),
which requires us to estimate the cooling function.
This, in turn, depends on the assumed mechanism of photon emission and on the
temperature and metallicity of the gas.
We use an Astrophysical Plasma Emission Code (APEC\footnote{\url{http://atomdb.org/}})
model for the X-ray emissivity, with the addition of a photoelectric absorption
(phabs\footnote{\href{http://heasarc.gsfc.nasa.gov/docs/xanadu/xspec/manual/XSmodelPhabs.html}{Xspec manual: phabs}})
from the foreground galactic gas.
For the latter, we adopt an equivalent column density of $3.04 \times 10^{20}$ cm$^{-2}$,
measured from the LAB Survey of Galactic HI \citep{Kalberla2005} in a cone of
radius $1\arcdeg$ centered on MACS~0416.
Finally, as the cooling function has only a weak dependence on the temperature
$T$ in the energy range considered in this work \citep{Ettori2000}, we assume a
constant gas temperature of $10.8$ keV and a metallicity of $0.24$, i.e., the
median values measured from the \emph{Chandra} data within a circle with radius
of $2 \arcmin$.
It is worth noticing that these temperature and metallicity values are extremely
close to those of $10.06$ keV and $0.24$, estimated within a larger
radius of $3\arcmin.75$ by \citet{Ogrean2015}.
From these values, we can compute the photon rate for the intracluster hot gas.

A uniform background of $0.805$~counts/pixel \citep[similar to the value used in][]{Ogrean2015}
is then added to the model; this has been measured in a  region of the image
sufficiently far away ($\approx 4\arcmin$) from the cluster emission.
We have checked that using a lower value for the background emission (i.e., $0.6$~counts/pixel), or leaving it as
a free parameter, does not change appreciably the value of the fitting figure of
merit and results in a difference in the cumulative projected gas mass of at maximum $4\%$ at $350$ kpc from the
cluster center.

To infer the values of the parameters of the model, we use the software
\emph{Sherpa}\footnote{\url{http://cxc.harvard.edu/sherpa}}. We adopt the Cash statistic $C$
\citep{Cash1979} as the likelihood function, as this is more appropriate than
a traditional Gaussian likelihood for the low counts of a Poisson distribution.
We restrict the fit to the inner circular region with a radius equal to
$40$ image pixels (i.e., approximately $840$~kpc) of
the surface brightness map and mask all point sources, found with the \emph{wavedetect}
algorithm.

To satisfactorily describe the X-ray surface brightness of MACS~0416, we adopt a model
that consists of three spherical dPIE components, with a resulting total surface
brightness proportional to the expression given in Equation (\ref{eq:dNPIE}).
Each of the three components has five free parameters: the
position of the center, $x_0$ and $y_0$, the central velocity dispersion,
$\sigma_0$, the core radius, $R_C$, and the truncation radius, $R_T$.
\begin{table}
\centering
\begin{tabular}{lrrr}

\hline
\hline
Parameter & Northeast 1 & Northeast 2 & Southwest \\
\hline

$x_0$ [$\arcsec$] & $-30$ & $-2$ & $29$ \\
$y_0$ [$\arcsec$] & $21$ & $0$ & $-50$ \\
$\sigma_0$ [km~s$^{-1}$] & $317$ & $201$ & $328$ \\
$R_{C}$ [$\arcsec$] & $34$ & $13$ & $35$ \\
$R_{T}$ [$\arcsec$] & $>5\times 10^3$ & $>750$ & $210$ \\
\hline

\end{tabular}
\caption{Best-fitting values of the parameters of the three-component dPIE (with
vanishing ellipticity) model of the X-ray surface brightness of MACS~0416.
Centers are relative to the North-East BCG, G1 (R.A. $=$ 04:16:09.154 and
decl. $= -$24:04:02.90).\label{tab:sbfit_best}}
\end{table}
Table \ref{tab:sbfit_best} presents the best-fitting values of the parameters of
the three dPIE components (the center coordinates are given with
respect to the northern BCG).
The final value of the adopted statistic is $6448$, which corresponds to a reduced value of
$1.31$, given the $4914$ degrees of freedom (dof) of the model.

\begin{figure*}
	\centering
	\includegraphics[width=\textwidth]{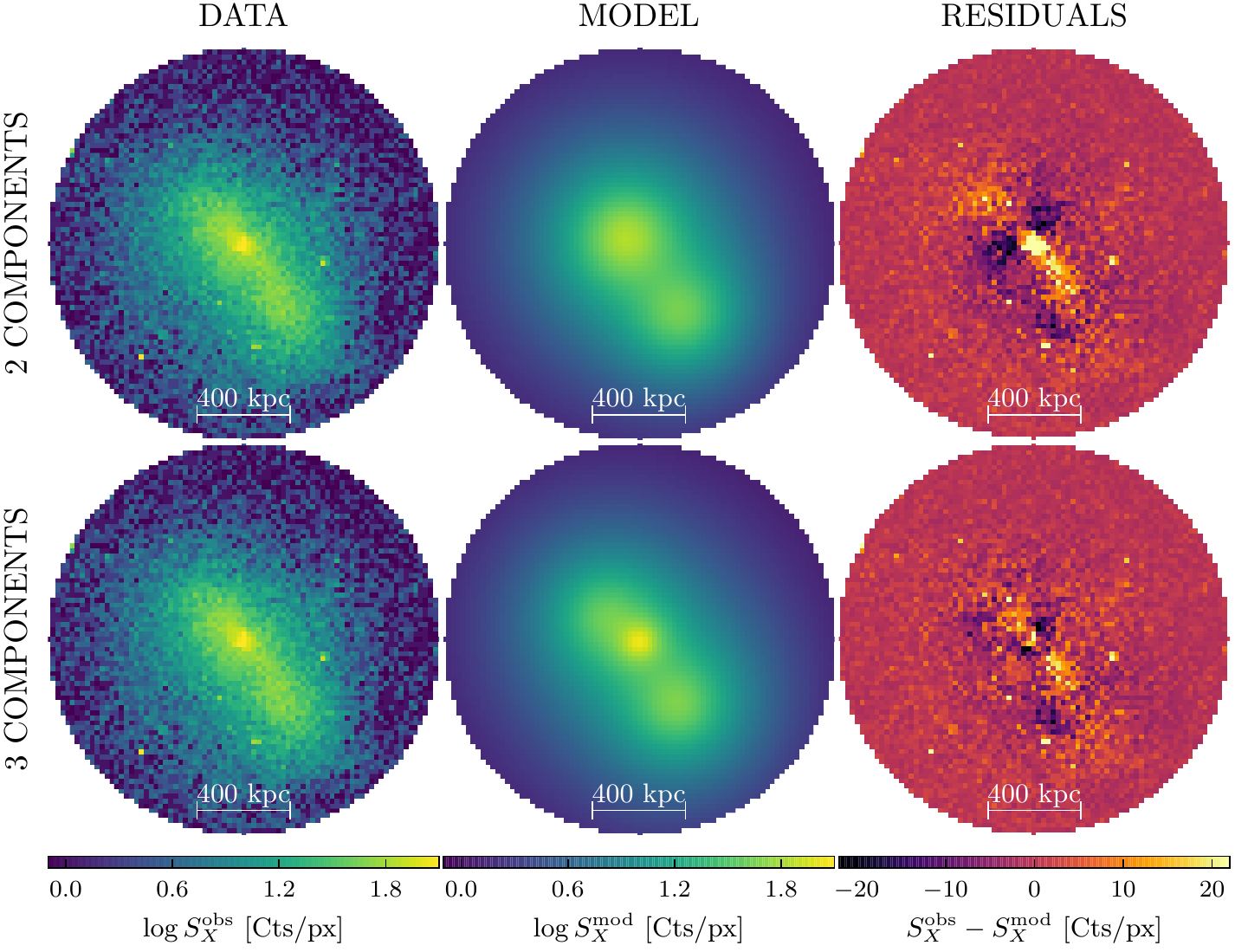}
	\caption{X-ray surface brightness (logarithmic scale) and residual maps of
	MACS~0416: observed counts (first column), best-fitting model (second column),
	and residuals (third column).
	The top and bottom rows show the maps for the two-component and
	three-component models, respectively.
	Each panel shows the circular region with a radius of $40$ image pixels (i.e., $\approx 840$
	kpc) used in the fitting procedure.
	The point sources are shown only for graphical reasons and have been masked
	out in the fitting procedure.}
	\label{fig:surface_brightness_data-model}
\end{figure*}

The hot gas distribution of the cluster is well represented by two diffuse components
(Northeast 1 and Southwest in Table \ref{tab:sbfit_best}) with values of central velocity dispersion of
approximately $320$~km~s$^{-1}$ and of core radius of about $180$~kpc.
The northeast clump requires an additional, more compact ($R_C \approx 70$~kpc),
component with a central velocity dispersion value of approximately $200$~km~s$^{-1}$.
All three dPIE components show very large values of $R_T$, extending beyond the
radius of the fitted region.
In passing, we note that for large values of $R_T$, a dPIE profile
becomes very similar to a $\beta$-model profile in the central regions.
The combination of the two northern components gives the total density for the main
northeast gas clump, and neither of them corresponds to the third dPIE
distribution needed in the lensing analysis (see \ref{sec:lens_model}).
The relative positions of the northern components are noteworthy: the compact
component is centered on the BCG, while the diffuse one is displaced by almost
$200$~kpc.
This offset creates the asymmetric emission that is seen in the
X-ray surface brightness in the first two columns of Figure \ref{fig:surface_brightness_data-model},
respectively, observations and models, respectively (see later a more detailed description of the figure).
Such asymmetry was observed already by \citet{Ogrean2015} and can be tentatively
interpreted as a tail formed as the subcluster approaches the first core passage
in a merging scenario.

Interestingly, our centers of the two main components are in moderate agreement with the
results of \citet{Jauzac2015}, but they find that the northeast distribution
has a core radius almost three times larger than the southwest one, while the
values of our core radii are very similar and approximately equal to $180$~kpc.
The lack of information on the fitting procedure implemented in \citet{Jauzac2015},
such as the size of the analyzed region, does not allow a more detailed comparison
between the two works.

The addition of a third component is necessary to accurately reproduce the compact emission
coming from the center of the northernmost clump, as it can be clearly seen in
Figure \ref{fig:surface_brightness_data-model}.
Here we show data, model, and residuals of a two- and three-spherical-component
model, in the top and bottom rows, respectively.
The first two columns represent the logarithm of the photon counts of the
observations and models, while the last one shows the residuals.
It is evident that the two-component model cannot fit well, at the same time,
the large-scale diffuse emission and the central compact peak.
A similar result has been found by \citet{Ogrean2015}, which used a double
$\beta$-model to fit the northeast subcluster.
A two-component model underpredicts the photon counts in the image inner regions, which
correspond to the position of G1, the northern BCG.
As a further test, we tried a model that consists only of two elliptical
components: the fit continues to fail in representing the bright central peak of
emission and the reduced statistic did not improve compared to the three
spherical mass distributions.

\begin{table}
\centering
\begin{tabular}{lrrrr}

\hline
\hline
	                & d.o.f.  & $C$     & AIC   & BIC  \\
\hline
2 spherical dPIE    & 4919    & 7391   & 7411  & 7475 \\
2 elliptical dPIE   & 4915    & 6817   & 6845  & 6936 \\
3 spherical dPIE    & 4914    & 6448   & 6478  & 6575 \\
\hline

\end{tabular}
\caption{Comparison between different models for the gas distribution: two spherical,
two elliptical and three spherical dPIE components. Columns show the degrees of
freedom (d.o.f.), the minimum value of the fitting Cash statistic $C$, the Akaike
information criterion (AIC) and the Bayesian Information Criterion (BIC).\label{tab:chis}}
\end{table}
We summarize the results of these three different models in Table \ref{tab:chis}, where
we show the degrees of freedom and the minimum value of the fitting Cash statistic $C$.
Furthermore, we include the values of the Akaike information criterion \citep[AIC,][]{Akaike1974} and
Bayesian Information Criterion \citep[BIC,][]{Schwarz1978}, two quantities that are often used for model comparison.
All of these criteria show that the three spherical component model is
preferable to a two component one, whether we include ellipticity or not.

\begin{figure}
	\centering
	\includegraphics[width=\columnwidth]{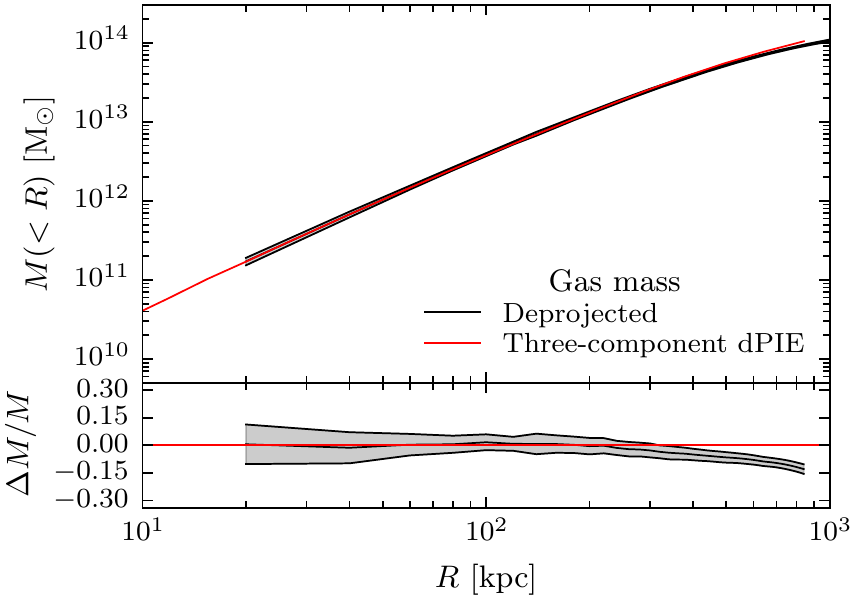}
	\caption{Top panel: cumulative projected gas mass profiles of MACS~0416.
	The red curve shows the best-fitting three-component dPIE model, while the
	central black curve shows the values obtained with a deprojection technique.
	The gray region, delimited by solid black curves, indicates the $\pm 1 \sigma$
	errors for the deprojected profile.
	Bottom panel: relative difference between the deprojected and
	three-component dPIE model gas mass profiles.
    The central curve shows the median values, while the gray region
	(delimited by solid black curves) shows the $\pm 1 \sigma$ errors derived
	only from the deprojected profile uncertainties.}
	\label{fig:gas_mass}
\end{figure}
As a final check, in Figure \ref{fig:gas_mass}, we have compared the cumulative projected gas mass profile of the
best-fitting three-component model with that obtained by directly deprojecting the gas.
The latter has been recovered through the geometrical deprojection
\citep[see, e.g.,][and references therein]{Ettori2013} of the azimuthally
averaged surface brightness profile that considers the entire X-ray emission,
as described in \citet[Appendix A]{Balestra2016}.
This 3D mass profile is then projected along the line of sight to estimate the
quantity shown in Fig.~\ref{fig:gas_mass}.
The top panel of the figure shows the radial profiles as computed
from the three-component dPIE model (red line) and from deprojecting the surface
brightness (black line with $1 \sigma$ errors shown as gray regions); the bottom panel, on the other hand,
gives the relative difference between the two mass measurements.
Errors in the bottom panel are derived only from the errors in the
deprojected profile; therefore, they represent a lower limit of the
uncertainties.
The agreement between the two different methods in the central region
($R<350$~kpc), which is the one of interest in the following strong-lensing
analysis, is remarkably good.

\section{Strong-lensing Analysis}\label{sec:lensing_analysis}
As done in Ca17, we combine the exquisite quality of the HFF images
\citep{Lotz2017} with the power of the MUSE integral-field spectrograph
\citep{Bacon2012} to reconstruct the total mass distribution of MACS~0416.
Due to the small differences with the work of Ca17, namely, the separate
treatment of two member galaxies (see Section \ref{sec:lens_model}),
we decided to rerun the reference model (hereafter REF).
The new model (hereafter GAS) presented in this work is the same as REF, but
with the hot gas included as a separate fixed mass component.

We briefly describe the multiple image and member galaxy catalogs we use in
this analysis; for more details we refer the reader to Ca17.
The multiple-image catalog contains $102$ images from $37$
systems spanning a range in redshift from $z\approx 0.94$ to $z\approx 6.15$ and
consists only of systems with secure identification \citep{Balestra2016,Caminha2016a}
and spectroscopic redshift.
The $193$ member galaxies have been selected on the basis of the available
photometric HST and spectroscopic VLT data; in detail, $144$ galaxies have a
spectroscopic redshift measurement, while the remaining ones have been chosen
based on their $n$-dimensional distance, in color space, from the locus of the
spectroscopically confirmed member galaxies (for more details, see Gr15).

The redshifts have been estimated from two sets of archival MUSE observations of
the northeast (program ID~094.A-0115B, PI: J.~Richard) and southwest (program
ID~094.A-0525(A), PI: F.~E.~Bauer) regions.

\subsection{Lens Mass Modeling}\label{sec:lens_model}
We distinguish the different mass components of the cluster into three main families:
diffuse main halos (mainly dark matter plus a few percent intracluster light),
member galaxies (with their respective dark matter halos) and hot gas (as discussed in Section
\ref{sec:xray_analysis}).

All components are described by dPIE profiles (see Equation
\ref{eq:dpie_3Ddens}), for which the convergence is
\begin{equation}
\kappa(x,y) = \frac{\sigma_0^2 R_T}{ 2 G \Sigma_{cr} \left( R_T - R_C \right) }
\left( \frac{1}{ \sqrt{ R_C^2 + R^2 } } - \frac{1}{ \sqrt{ R_T^2 + R^2 } } \right),
\label{eq:dpie_kappa}
\end{equation}
where $\sigma_0$ is the central velocity dispersion, $R_C$ is the core radius,
$R_T$ is the truncation radius, $\Sigma_{cr}$ is the critical surface density,
and $R$ is the radial distance from the mass center.
Including a possible elongation term on the plane of the sky, the definition of
$R$ becomes $R^2 = x^2 ( 1 + \epsilon )^{-2} + y^2 ( 1 - \epsilon )^{-2}$,
with the ellipticity $\epsilon$ defined as $\epsilon \equiv ( 1 - q )/( 1 + q )$
and $q$ being the minor-to-major-axis ratio.

Given the complex structure of MACS~0416, visible for example in its luminosity
distribution, we adopt two dPIE components for the diffuse halos.
We fix their truncation radii to infinity, effectively making them equivalent to 
pseudo-isothermal elliptical mass distribution \citep[hereafter PIEMD;][]{Kassiola1993} profiles.
Each of these two components has six free parameters: the center coordinates, $x_h$ and
$y_h$, the ellipticity and position angle, $\epsilon_h$ and $\theta_h$, the core
radius, $R_{C,h}$, and the central velocity dispersion, $\sigma_{0,h}$.
As explained in Ca17, the addition of a third diffuse halo is
required to reduce the offset between the positions of observed and model-predicted multiple
images in the northeast region of the northern BCG, G1.
This halo is assumed to be spherical; therefore, it is described by only four
parameters: $x_{h3}$, $y_{h3}$, $R_{C,h3}$, and $\sigma_{0,h3}$.

Each member galaxy is modeled with a spherical dPIE profile with vanishing core
radius and center fixed at the position of the galaxy luminosity center.
To reduce the number of free parameters, we scale the values of $\sigma_{0,i}$
and $R_{T,i}$ of each galaxy depending on its luminosity $L_i$ (in the HST/WFC3
filter F160W).
We refer to these values as $\sigma_{0,g}$ and $R_{T,g}$, the velocity dispersion and truncation
radius of the reference galaxy, G1, with luminosity $L_{g}$
($mag_{\text{F160W}} = 17.02$),
\begin{equation}
\sigma_{0,i} = \sigma_{0,g} \left( \frac{L_i}{L_g} \right)^{0.35} \;\; \text{and}  \;\;
R_{T,i} = R_{T,g} \left( \frac{L_i}{L_g} \right)^{0.5}.
\end{equation}
These scaling relations have been chosen as they reproduce the tilt of the
fundamental plane \citep{Faber1987,Bender1992} observed in early-type galaxies.
Additionally, they have been shown to describe accurately the total mass
properties of member galaxies in MACS~0416 (Gr15, Ca17), MACS J1149.5+2223
\citep{Grillo2016}, and RXC J2248.7$-$4431 \citep{Caminha2016a}.
We do not treat separately the two member galaxies mainly responsible for the
appearance of the multiple images of family 14, contrary to what has been done by Ca17.
We model these two galaxies using the same scaling relations adopted for the
other member galaxies.

We include an additional galaxy halo at the location of a foreground
galaxy in the southwest region of the cluster (R.A. $=$ 04:16:06.82 and
decl. $= -$24:05:08.4).
Given that this galaxy does not belong to the cluster ($z=0.112$), its
$\sigma_0$ and $R_T$ should be considered only as effective parameters.

In the GAS model, we include a component for the hot gas distribution, as derived
from the analysis of the X-ray surface brightness presented in Section \ref{sec:xray_analysis}.
We keep this component fixed, when fitting the multiple image positions.
The inclusion as a fixed component is justified by the smaller set of
assumptions required to derive the gas density profile from
the X-ray surface brightness.
Besides, the statistical errors on the hot gas mass profile are smaller than
those typically associated with the other cluster mass components.

We use the software \emph{lenstool} \citep{Jullo2007} to infer the best-fitting
values of the parameters of the total mass models of MACS~0416, using the
positions of multiple images as observables.
Furthermore, we adopt uniform priors for all model parameters.

In summary, we use two descriptions for the cluster total mass: a reference model (REF), where the
diffuse component includes the hot gas, and the new one (GAS), where the hot gas
distribution is fixed to the result of an X-ray surface brightness analysis.
Both of them have the same number of free parameters, as the separate hot gas
mass density is not optimized in the lensing modeling.

For each model, we initially adopt an error on the position of the multiple images
of $0.5\arcsec$, the same value as in Ca17 and close to the theoretical prediction
by \citet{Jullo2010}.
The resulting best-fitting models have a reduced $\chi^2$ larger than unity
($1.2$ and $1.3$ for REF and GAS, respectively).
In order to get realistic uncertainties for the model parameters, the $\chi^2$
should be comparable with the number of degrees of freedom (110).
Therefore, we rerun the MCMC analysis using an image positional error of $0.58\arcsec$.
This value has been obtained by requiring that the reduced $\chi^2$ of the
best-fitting models is approximately $1$.
By increasing the uncertainty of the image positions, we take into
account unknown factors, such as line-of-sight mass structures or small
dark matter clumps in the cluster, which affect the observed positions of the images.
Hereafter, we will refer only to the second runs, with $0.58\arcsec$ positional errors, when
presenting the results.
We run the MCMC analysis until convergence, resulting in a total of more than $1.1 \times 10^5$
points that sample the posterior probability distribution of the model parameters.

\section{Results}
\label{sec:results}
In this section, we present the results of the two mass models we have obtained
for MACS~0416, namely, the reference model (REF) and the model where
the hot gas component has been included separately (GAS).

\renewcommand{\arraystretch}{1.3}

\begin{table*}
\centering
\begin{tabular}{l*{4}{c}*{4}{c}r}

\hline
\hline
 & \multicolumn{4}{c}{REF} & \multicolumn{4}{c}{GAS} & \\
\hline
 & Median & 68\% CL & 95\% CL & 99.7\% CL & Median & 68\% CL & 95\% CL & 99.7\% CL & \\
\hline

$x_{h1}$ [$\arcsec$] & $-2.0$ & $^{ +1.0 } _{ -1.0 }$ & $^{ +1.8 } _{ -1.7 }$ & $^{ +2.8 } _{ -2.3 }$ & $-2.4$ & $^{ +1.0 } _{ -0.8 }$ & $^{ +2.3 } _{ -1.5 }$ & $^{ +3.4 } _{ -2.2 }$ & $x_{h1}$ [$\arcsec$] \\
$y_{h1}$ [$\arcsec$] & $1.4$ & $^{ +0.7 } _{ -0.7 }$ & $^{ +1.2 } _{ -1.5 }$ & $^{ +1.6 } _{ -2.2 }$ & $1.7$ & $^{ +0.5 } _{ -0.7 }$ & $^{ +0.9 } _{ -1.7 }$ & $^{ +1.3 } _{ -2.7 }$ & $y_{h1}$ [$\arcsec$] \\
$\epsilon_{h1}$ & $0.84$ & $^{ +0.02 } _{ -0.06 }$ & $^{ +0.04 } _{ -0.11 }$ & $^{ +0.05 } _{ -0.14 }$ & $0.85$ & $^{ +0.02 } _{ -0.02 }$ & $^{ +0.03 } _{ -0.08 }$ & $^{ +0.04 } _{ -0.12 }$ & $\epsilon_{h1}$ \\
$\theta_{h1}$ [deg] & $144.7$ & $^{ +1.2 } _{ -1.1 }$ & $^{ +2.8 } _{ -2.5 }$ & $^{ +4.2 } _{ -4.3 }$ & $145.1$ & $^{ +0.9 } _{ -0.9 }$ & $^{ +2.1 } _{ -2.0 }$ & $^{ +3.9 } _{ -3.6 }$ & $\theta_{h1}$ [deg] \\
$R_{C,h1}$ [$\arcsec$] & $6.7$ & $^{ +0.9 } _{ -0.9 }$ & $^{ +1.7 } _{ -1.8 }$ & $^{ +2.5 } _{ -2.8 }$ & $6.8$ & $^{ +0.8 } _{ -1.0 }$ & $^{ +1.5 } _{ -1.9 }$ & $^{ +2.4 } _{ -2.7 }$ & $R_{C,h1}$ [$\arcsec$] \\
$\sigma_{0,h1}$ [km s$^{-1}$] & $713$ & $^{ +32 } _{ -34 }$ & $^{ +60 } _{ -70 }$ & $^{ +90 } _{ -120 }$ & $708$ & $^{ +26 } _{ -29 }$ & $^{ +48 } _{ -67 }$ & $^{ +71 } _{ -102 }$ & $\sigma_{0,h1}$ [km s$^{-1}$] \\
$x_{h2}$ [$\arcsec$] & $20.1$ & $^{ +0.4 } _{ -0.5 }$ & $^{ +1.1 } _{ -1.0 }$ & $^{ +1.7 } _{ -1.4 }$ & $20.0$ & $^{ +0.4 } _{ -0.4 }$ & $^{ +0.9 } _{ -0.9 }$ & $^{ +1.6 } _{ -1.3 }$ & $x_{h2}$ [$\arcsec$] \\
$y_{h2}$ [$\arcsec$] & $-37.1$ & $^{ +0.8 } _{ -0.8 }$ & $^{ +1.6 } _{ -1.7 }$ & $^{ +2.3 } _{ -2.7 }$ & $-37.0$ & $^{ +0.7 } _{ -0.7 }$ & $^{ +1.5 } _{ -1.5 }$ & $^{ +2.1 } _{ -2.6 }$ & $y_{h2}$ [$\arcsec$] \\
$\epsilon_{h2}$ & $0.76$ & $^{ +0.02 } _{ -0.02 }$ & $^{ +0.03 } _{ -0.03 }$ & $^{ +0.05 } _{ -0.07 }$ & $0.77$ & $^{ +0.02 } _{ -0.01 }$ & $^{ +0.03 } _{ -0.03 }$ & $^{ +0.04 } _{ -0.05 }$ & $\epsilon_{h2}$ \\
$\theta_{h2}$ [deg] & $125.8$ & $^{ +0.5 } _{ -0.5 }$ & $^{ +1.0 } _{ -1.0 }$ & $^{ +1.6 } _{ -1.5 }$ & $125.9$ & $^{ +0.4 } _{ -0.4 }$ & $^{ +0.9 } _{ -0.8 }$ & $^{ +1.4 } _{ -1.3 }$ & $\theta_{h2}$ [deg] \\
$R_{C,h2}$ [$\arcsec$] & $13.2$ & $^{ +0.9 } _{ -0.9 }$ & $^{ +1.7 } _{ -1.6 }$ & $^{ +2.5 } _{ -2.3 }$ & $12.6$ & $^{ +0.7 } _{ -0.7 }$ & $^{ +1.4 } _{ -1.4 }$ & $^{ +2.2 } _{ -2.2 }$ & $R_{C,h2}$ [$\arcsec$] \\
$\sigma_{0,h2}$ [km s$^{-1}$] & $1103$ & $^{ +22 } _{ -22 }$ & $^{ +43 } _{ -45 }$ & $^{ +73 } _{ -81 }$ & $1065$ & $^{ +19 } _{ -20 }$ & $^{ +39 } _{ -38 }$ & $^{ +62 } _{ -60 }$ & $\sigma_{0,h2}$ [km s$^{-1}$] \\
$x_{h3}$ [$\arcsec$] & $-34.3$ & $^{ +1.2 } _{ -1.5 }$ & $^{ +2.2 } _{ -3.7 }$ & $^{ +3.4 } _{ -6.8 }$ & $-34.3$ & $^{ +1.0 } _{ -1.3 }$ & $^{ +2.0 } _{ -3.4 }$ & $^{ +3.1 } _{ -5.8 }$ & $x_{h3}$ [$\arcsec$] \\
$y_{h3}$ [$\arcsec$] & $8.7$ & $^{ +3.3 } _{ -1.3 }$ & $^{ +4.9 } _{ -2.0 }$ & $^{ +6.9 } _{ -2.8 }$ & $8.1$ & $^{ +1.6 } _{ -0.8 }$ & $^{ +5.0 } _{ -1.4 }$ & $^{ +6.8 } _{ -2.1 }$ & $y_{h3}$ [$\arcsec$] \\
$R_{C,h3}$ [$\arcsec$] & $7.5$ & $^{ +2.4 } _{ -2.7 }$ & $^{ +4.9 } _{ -5.2 }$ & $^{ +8.0 } _{ -7.1 }$ & $4.6$ & $^{ +3.0 } _{ -2.4 }$ & $^{ +5.5 } _{ -4.0 }$ & $^{ +7.9 } _{ -4.5 }$ & $R_{C,h3}$ [$\arcsec$] \\
$\sigma_{0,h3}$ [km s$^{-1}$] & $435$ & $^{ +59 } _{ -62 }$ & $^{ +125 } _{ -117 }$ & $^{ +192 } _{ -156 }$ & $351$ & $^{ +64 } _{ -51 }$ & $^{ +131 } _{ -87 }$ & $^{ +191 } _{ -112 }$ & $\sigma_{0,h3}$ [km s$^{-1}$] \\
$R_{T,g}$ [$\arcsec$] & $7.8$ & $^{ +4.3 } _{ -2.4 }$ & $^{ +11.9 } _{ -3.5 }$ & $^{ +37.0 } _{ -4.3 }$ & $7.7$ & $^{ +3.6 } _{ -2.0 }$ & $^{ +10.5 } _{ -3.7 }$ & $^{ +14.9 } _{ -4.5 }$ & $R_{T,g}$ [$\arcsec$] \\
$\sigma_{0,g}$ [km s$^{-1}$] & $321$ & $^{ +27 } _{ -77 }$ & $^{ +46 } _{ -102 }$ & $^{ +61 } _{ -136 }$ & $317$ & $^{ +17 } _{ -72 }$ & $^{ +44 } _{ -103 }$ & $^{ +62 } _{ -128 }$ & $\sigma_{0,g}$ [km s$^{-1}$] \\
\hline

\end{tabular}
\caption{\label{tab:medians}Values of the parameters of the lens models of MACS~0416.
    Median values and confidence level (CL) uncertainties are given for the two models presented in the
	paper. Centers are relative to the North-East BCG, G1 (R.A. $=$ 04:16:09.154 and decl.: $= -$24:04:02.90). The angles
	$\theta_{h1}$ and $\theta_{h2}$ are measured counterclockwise from the West
	axis.}
\end{table*}
\renewcommand{\arraystretch}{1.}

The values of the parameters inferred for the two models are shown in Table
\ref{tab:medians}, where we quote the median values and the $68\%$, $95\%$ and $99.7\%$
confidence level (CL) intervals.
The positions of the centers are given relatively to the northern BCG, G1.
The best-fit $\chi^2$ values (logarithmic Bayesian evidence)  are $99.5$
($-202.3$) and $105.5$ ($-201.3$) for the REF and GAS model,
respectively, corresponding to root mean square values of  $0.57\arcsec$ and
$0.59\arcsec$ (median values $0.40\arcsec$ and $0.41\arcsec$) for the offset
between the observed and model-predicted positions of the multiple images.
Given that both models have the same number of degrees of freedom (110), it is 
remarkable that we obtain similar $\chi^2$ values when we fix some per cent
(about $10\%$, as estimated below) of the cluster total mass in the hot gas
component.

In the case of the Bullet cluster, \citet{Paraficz2016} found that the model
with a separate hot gas component is strongly preferred, mostly due to the large offset
between the gas and dark matter components.
Our new approach allows us to more accurately characterize the collisional and collisionless components, even in
less extreme merging conditions.

\begin{figure}
	\centering
    	\includegraphics{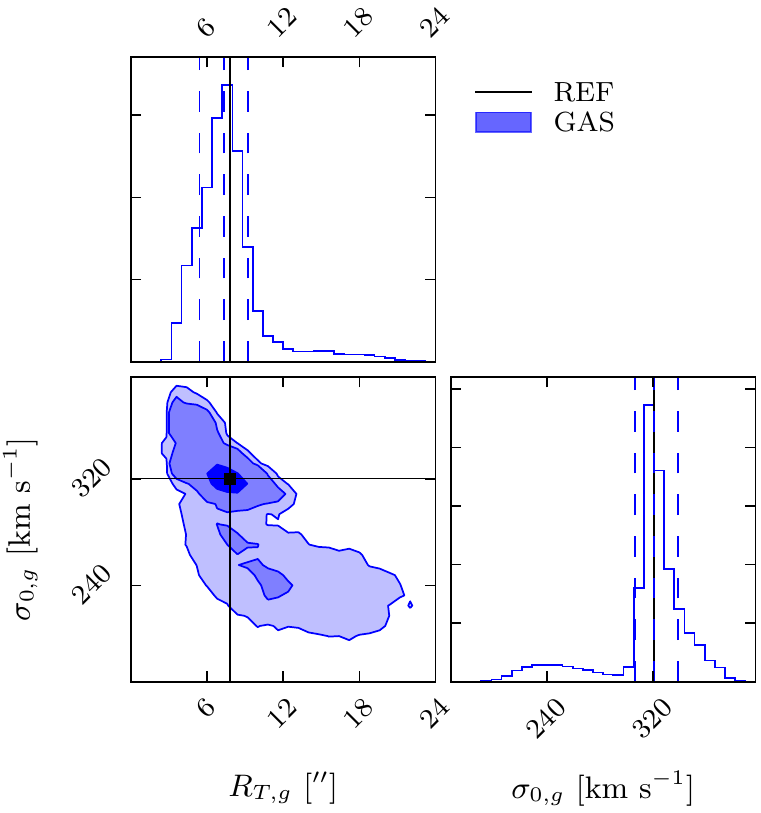}
	\caption{Posterior distributions of the values of the North-East BCG (G1) parameters (entering in Equation \ref{eq:dpie_kappa})
	derived from the lensing model with the hot gas component included separately (i.e., GAS model).
	Blue contours contain $39.3\%$,  $86.5\%$ and $98.9\%$ of the samples,
	while blue dashed lines in the 1D histograms show the $16^\text{th}$, $50^\text{th}$,
	and $84^\text{th}$ percentiles. Black solid lines mark the median values of the
	reference model (i.e., REF model).}
	\label{fig:corner_gas_comp_Pot0}
\end{figure}
For the GAS model, we show the posterior probability
distributions of the parameters of the three main halos (see Appendix \ref{sec:posteriors}) and of the
member galaxy scaling relations (see Figure \ref{fig:corner_gas_comp_Pot0}).
Here the blue contours contain the  $39.3\%$,  $86.5\%$ and $98.9\%$ of the
samples, which correspond to  the $1\sigma$, $2\sigma$ and $3\sigma$ of a 2D Gaussian
distribution; the black solid lines are the median values of the reference model.
In the 1D histograms, vertical blue dashed lines coincide with $16^\text{th}$,
$50^\text{th}$, and $84^\text{th}$ percentiles.
Out of the correlation plots not shown here, there are no strong degeneracies
among different halo parameters.
Only weak correlations are visible between the parameters of the northeast and
third halos, which are to be expected given their close distance in projection.

Interestingly, the MCMC chains converge to values that are very similar to those
presented in Ca17, with the exception of $R_{T,g}$ and $\sigma_{0,g}$; however,
these are still consistent within $1\sigma$ uncertainties.
While our reference model seems to favor more compact member galaxies with
higher central velocity dispersions, Ca17 found a median value of $\sigma_{0,g}$
that corresponds with the secondary peak ($\sigma_{0,g} \approx 240$ km s$^{-1}$) in our marginalized distribution shown in Figure
\ref{fig:corner_gas_comp_Pot0}.

The inclusion of a separate hot gas component does not change substantially the
inferred properties of the diffuse halo components (all the values of their
parameters are within $3\sigma$ with respect to those of the reference model).
As expected, the velocity dispersion and core radius values of the diffuse halos change the most, as
these parameters are proportional to the square root of the mass of the
corresponding component.
This follows from the fact that the gas mass is now modeled separately and not
included in the diffuse halos, as in the reference model.

\begin{figure}
	\centering
	\includegraphics[width=\columnwidth]{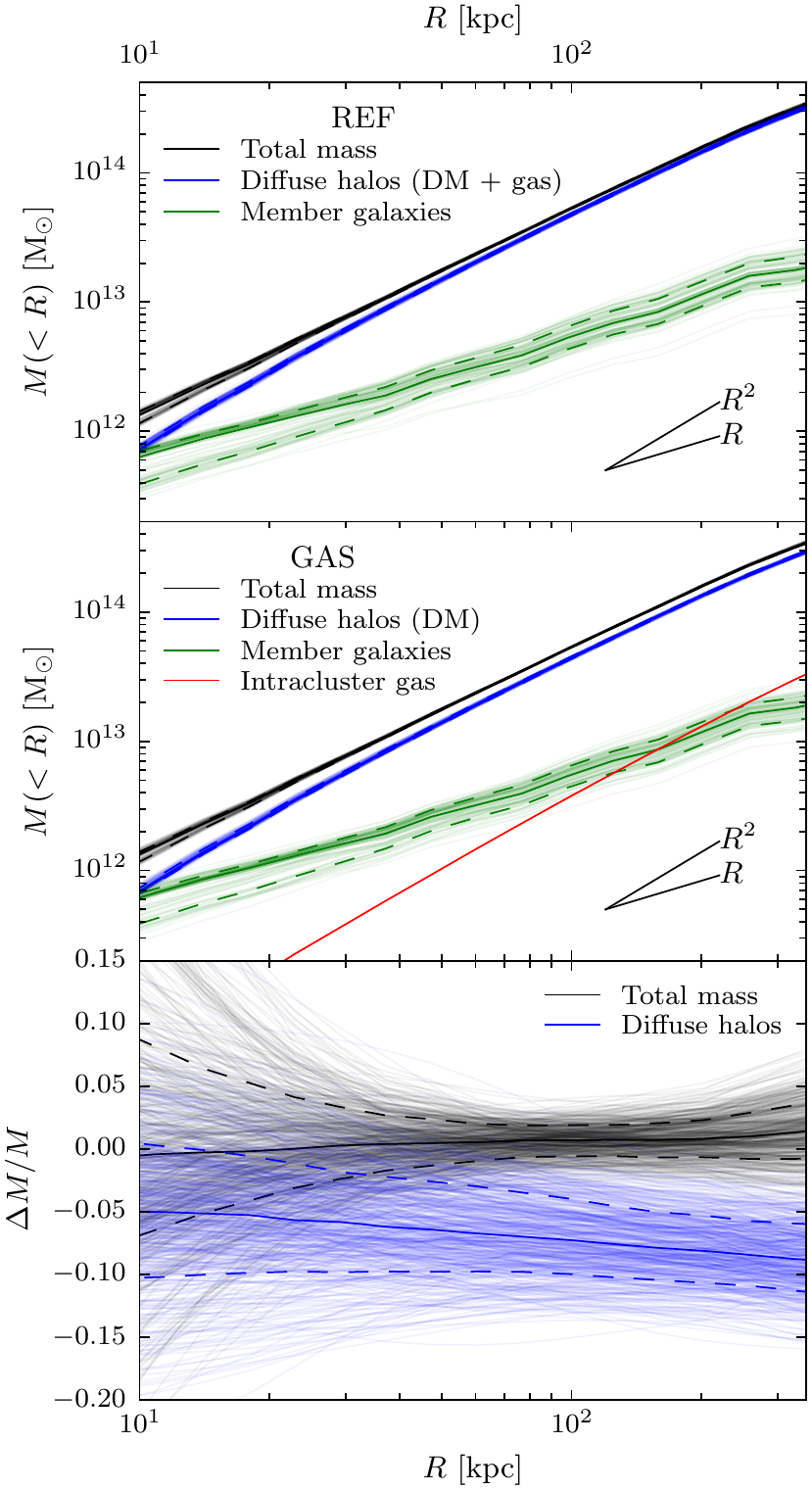}
	\caption{Radial profiles of cumulative projected mass of MACS~0416 for the reference
	(top panel) and separate gas (middle panel) models. Black, blue, green, and
	red curves represent the total, diffuse halos, galaxy members, and
	hot intracluster gas profiles, respectively.
	Thin lines show a subsample of the models in the final MCMC chains, while solid and dashed
	lines give the median and $16^\text{th}\,$-$\,84^\text{th}$ percentiles.
    We indicate the values of the central logarithmic slope corresponding to a
    cored and a singular isothermal sphere surface mass distribution, respectively $2$ and $1$, respectively.
	The bottom panel shows the relative difference between the same components
	of the two models: in black the total and in blue the diffuse halo mass.}
	\label{fig:enclosed_mass}
\end{figure}
We estimate the cumulative projected mass profile of the various components:
total, diffuse halos (mostly dark matter), member galaxies, and hot gas.
These are shown, respectively, in black, blue, green, and red in the top and
middle panels of Figure \ref{fig:enclosed_mass}.
A subsample of the MCMC chains is shown as thin lines, and their
median and $16^\text{th}\,$-$\,84^\text{th}$ percentiles are shown with solid
and dashed lines.
The top panel corresponds to the REF model, while the middle one corresponds to the GAS model.
Noticeably, the total and member galaxy mass profiles are very similar in the
two models.
This agreement confirms the above statement that the hot gas mass component is
essentially subtracted from that of the main halos, not affecting the total or
the member galaxy mass estimates.
The decrease of mass in the diffuse halos is more evident in the bottom panel of Figure
\ref{fig:enclosed_mass}, where the relative difference in enclosed mass between
the two models is shown for the total mass (black) and the diffuse halos (blue).
This plot simply shows the ratios of the two black and two blue curves of the
panels above.
We exclude from the figure a comparison of the member galaxy component, as the
error in the mass ratio is too large
to give any useful information.
As noted before, the total mass measurements of the two models are consistent
within the errors, while the mass in the diffuse halo component decreases,
once the hot gas is treated separately.
This difference is significant at more than $3\sigma$ above $100$~kpc and
we find approximately $10\%$ less mass (at a radial distance of approximately $350$~kpc
from the main BCG) in the diffuse halos in our refined model.

Interestingly, looking at the middle panel of Figure \ref{fig:enclosed_mass},
the mass in the diffuse halo component is almost a constant fraction of the total
mass, outside the region where the BCG contribution might still be very relevant
($R > 40$~kpc).
\begin{figure}
	\centering
	\includegraphics[width=\columnwidth]{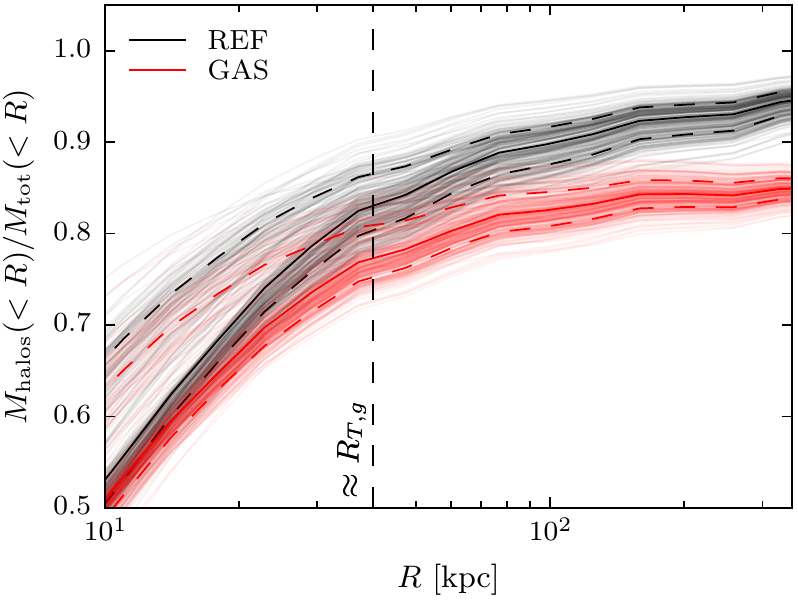}
	\caption{Radial profiles of the fraction of cumulative projected total mass
	in the diffuse halo component (mostly DM). Black curves show the halos that include
	both the dark matter and hot intracluster gas components (i.e., REF model),
	while the red curves represent the dark-matter-only component (i.e., GAS model).
	Thin lines show a subsample of the models in the final MCMC chains, while solid and dashed
	lines give the median and $16^\text{th}\,$-$\,84^\text{th}$ percentiles, respectively.
    The vertical dashed line shows approximately the truncation radius of the BCG G1.
	}
	\label{fig:dm_fraction}
\end{figure}
To better quantify this feature, in Figure \ref{fig:dm_fraction} we plot the
cumulative projected diffuse halo over total mass profile for the REF (black)
and GAS (red) models.
The central solid lines mark the median values, and the colored regions show the
$1\sigma$, $2\sigma$ and $3\sigma$ confidence regions, similarly to the previous plot.
Having removed the hot gas component from the diffuse halos, the red profile
describes more realistically than the black one the cluster dark matter over
total mass fraction.
Moving towards the center of the cluster, the fraction of dark matter varies
slowly and begins to decrease noticeably only around the BCG truncation radius
$R_{T,g} \approx 40$ kpc (marked with a vertical dashed line).

\begin{figure*}
	\centering
	\includegraphics[width=\textwidth]{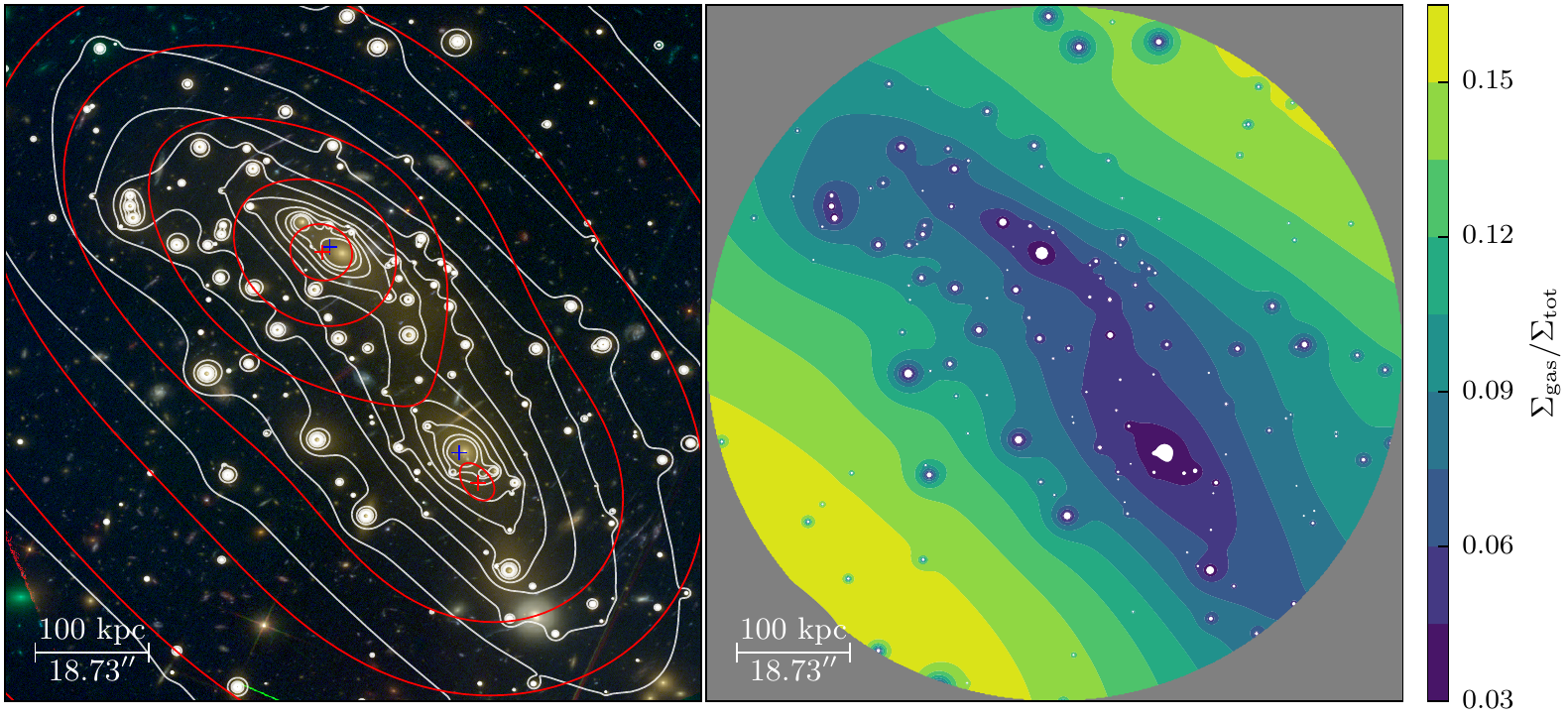}
	\caption{Left: total (white) and hot gas (red) surface mass density isocontours
	overlaid on a color-composite HST WCF3+ACS image of MACS~0416 \citep[seven filters from the Hubble
	Frontier Fields data; see][]{Caminha2017}. Total mass isodensities have a
	linear step of $2.5 \times 10^8$~M$_\odot$~kpc$^{-2}$ and go from $3.5 \times 10^8$ to
    $2.85 \times 10^9$~M$_\odot$~kpc$^{-2}$. Hot gas isodensities go from
    $4.5 \times 10^7$ to $1.35 \times 10^8$~M$_\odot$~kpc$^{-2}$ with a linear step of
    $1.5 \times 10^7$~M$_\odot$~kpc$^{-2}$. The peaks of the surface mass density
    maps are shown with blue and red plus signs, for the diffuse dark matter and intracluster
    hot gas, respectively. Right: local gas over total mass fraction map, derived
    from the surface mass density maps. White circles show the member galaxies,
    with each circle area proportional to the galaxy luminosity.}
	\label{fig:massmaps}
\end{figure*}
Moreover, we provide maps of the total and gas surface mass densities (left panel of
Figure \ref{fig:massmaps}), derived from our refined best-fitting model GAS, which we
then use to obtain a map of the local gas-to-total mass fraction (right panel).
The white curves, overlaid on top of the HFF color image, are isodensity contours of
the total mass, from $3.5 \times 10^8$ to
$2.85 \times 10^9$ M$_\odot/$kpc$^2$, with a linear step of $2.5 \times 10^8$
M$_\odot/$kpc$^2$; the red ones show the hot gas component, from
$4.5 \times 10^7$ to $1.35 \times 10^8$ M$_\odot/$kpc$^2$, with a linear step of
$1.5 \times 10^7$ M$_\odot/$kpc$^2$.
The blue and red plus signs mark the location of the maximum values in the dark
matter and intracluster hot gas surface mass density maps.
The filled contours shown in the right panel trace the gas-to-total mass
fraction computed from the maps of the gas and total surface mass densities.
The white circles locate the positions of the member galaxies, and the circle area is
proportional to the $H$-band luminosity of the galaxy they represent.
We mask (in gray) the outer regions of the cluster, where there is no lensing
information from observed multiple images, thus making the cluster mass
reconstruction here less accurate.
One of the reasons for the separate inclusion of the hot gas component in cluster lensing analyses is
evident from these two plots: dark matter and intracluster gas are distributed
slightly differently, the former appearing more elongated than the latter.

For the two main subclusters, we have computed the distance between the density
peaks of each component and the closest BCG.
In the northeast sector, these correspond to approximately $2\arcsec$ (peak
located at R.A. $=$ 4:16:09.276 and decl: $= -$24:04:01.87) and $3\arcsec$ (peak
located at R.A. $=$ 4:16:09.373 and decl: $= -$24:04:02.73), for the dark matter and
hot intracluster gas densities, respectively.
In the southwest subcluster, the densities of the two components peak at distances of
approximately $1\arcsec$ (peak located at R.A. $=$ 4:16:07.718 and 
decl: $= -$24:04:39.04) and $6\arcsec$ (peak located at R.A. $=$ 4:16:07.497 and
decl: $= -$24:04:44.49).
Given statistical and systematic uncertainties of a few arcseconds in the
position of the peaks, the only evidence for an offset is for the gas component
in the southwest (about $2\sigma$-$3\sigma$).
This offset is smaller than the one found in \citet{Jauzac2015}, and the position
of the hot gas density peak is consistent with the uncertainty region in the
X-ray surface brightness peak presented in \citet{Ogrean2015}; we refer to Gr15
and \citet{Balestra2016} for further details.

A difference in the centers of the dark matter and hot gas mass distributions is
another effect traditionally not included in strong-lensing models.
As mentioned previously \citep[see Section \ref{sec:intro} and][]{Markevitch2004},
it is in principle possible to determine the cross section of SIDM
from the offset between the member galaxy and dark matter distributions;
therefore it is extremely important to have an accurate measurement of the
center of the dark matter component, without the bias introduced by the intracluster
gas, which lags behind in merging events owing to its collisional behavior.

Naturally, possible differences between the distributions of dark matter and hot
gas are reflected into the gas over total mass fraction map.
For instance, the larger core of the hot gas mass component creates two peaks in
the map at the location of the two BCGs.

\begin{figure}
	\centering
	\includegraphics[width=\columnwidth]{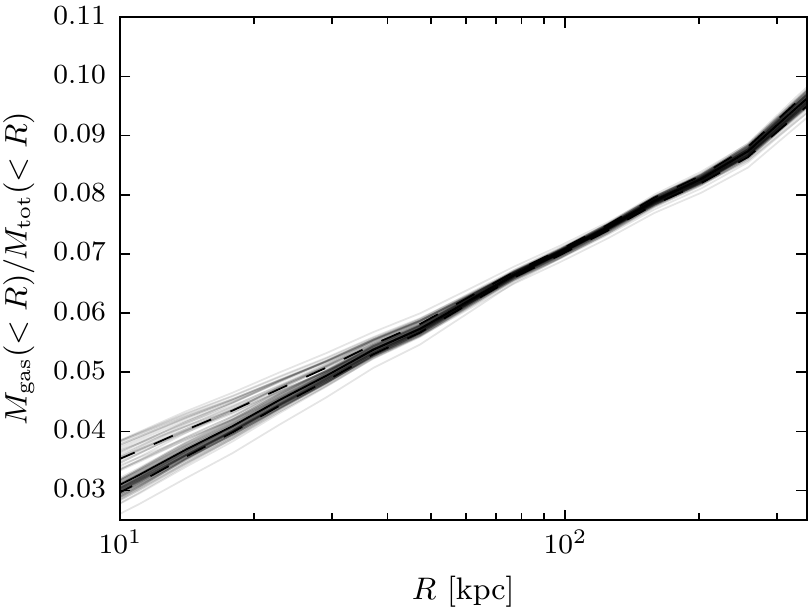}
	\caption{Cumulative projected gas over total mass fraction computed from the
	cumulative 2D mass profiles. Thin lines show
	a subsample of the MCMC models, while solid and dashed lines give the
	median and $16^\text{th}\,$-$\,84^\text{th}$ percentiles, respectively.}
	\label{fig:gas_fraction}
\end{figure}
Finally, we compute the cumulative projected gas over total mass fraction as a
function of the distance from G1, shown in Figure \ref{fig:gas_fraction}.
As in Figure \ref{fig:enclosed_mass}, solid and dashed lines show
the median and $16^\text{th}\,$-$\,84^\text{th}$ percentiles, while the thin
lines are obtained from a subsample of the models extracted from the
MCMC chain.
As we assume a fixed gas profile, the errors shown here underestimate the true
uncertainties in the gas fraction.
Clearly, gravitational lensing can only provide two-dimensional information
about the total mass of a lens, and a deprojection of the mass distribution in a
merging cluster, like MACS~0416, is not a trivial task.
This would require several assumptions about the symmetry of the system.
Hence, we decide to use our 2D mass densities in the computation of the gas fraction.
The projected gas over total mass fraction within an aperture of $350$~kpc is
approximately $10\%$.
A direct comparison with the results by \citet{Jauzac2015} is not possible
because the (de)projection method used to compute the gas over total
mass fraction, shown in their Figure 11, is not fully described.

\begin{figure}
	\centering
	\includegraphics[width=\columnwidth]{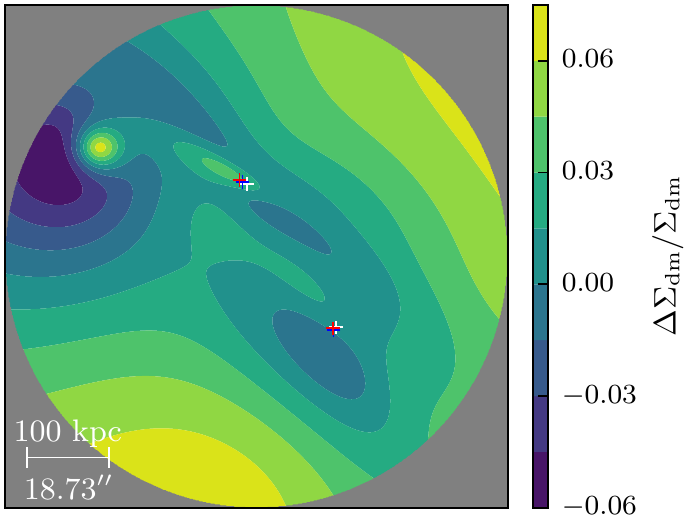}
	\caption{Relative difference between the proposed method and an a-posteriori
	analysis in the diffuse dark-matter surface mass density.
	Crosses mark the BCGs (white) and the peaks of the diffuse dark-matter distribution
	of the GAS (red) and POST (blue) models, respectively.}
	\label{fig:ratio_dm_posteriori}
\end{figure}
In passing, we mention that we have also subtracted the hot gas surface mass
density reconstructed in Section \ref{sec:xray_analysis} from the diffuse halo surface mass density of the
best-fit REF model, to mimic the results one would obtain in a posteriori analyses.
Figure \ref{fig:ratio_dm_posteriori} shows the relative difference between the
best-fit models of the proposed method (GAS) and an a posteriori analysis (POST)
in the estimate of the diffuse dark-matter-only surface mass density.
The white plus signs show the position of the BCGs, G1 and G2, and the peaks of
the diffuse dark matter distribution of the GAS and POST models are marked with red and
blue plus signs, respectively.
The differences in the dark matter surface mass density are quite noticeable in the North-East
region, around the third halo, but are overall small (less than $6\%$).
Similarly, the position of the peak of the northeast component changes
slightly, resulting in a difference of approximately $1\arcsec$ between the GAS and
POST models.
Although the dark matter surface mass density maps of the two
approaches in MACS~0416 are very similar, the advantage of the method we adopt
in this work is the direct measurement of the parameter values of the dark-matter-only distribution,
unavailable in a traditional analysis.

\section{Conclusions}
\label{sec:conclusions}
In this paper, we have presented a novel approach to include the hot gas
component in strong gravitational lensing analyses.
The method starts from a separate modeling of the X-ray surface brightness to derive
the distribution of the intracluster gas, which is then used as an additional
fixed mass component to the lensing fit. 
By doing so, we can disentangle the cluster hot gas from the diffuse main halo, thus
tracing more accurately the non collisional mass component, i.e. mainly dark
matter.
We have applied this method to the HFF merging cluster MACS J0416.1$-$2403, which was the
target of several recent spectroscopic campaigns \citep{Grillo2015,Balestra2016,Hoag2016,Caminha2017}.
We have fitted the observed positions of a large set of spectroscopically
confirmed multiple images with two lensing models: one adopting
our new technique (GAS) and one following a traditional modeling of the
intracluster gas included in the main diffuse halos (REF).

The main results of the work can be summarized as follows:
\begin{itemize}
\item We have provided an analytical solution for the X-ray surface brightness
emission produced by multiple dPIE mass density distributions.
This profile is commonly used in the lensing community and therefore readily
available ``out of the box'' in most gravitational lensing softwares, as opposed
to the $\beta$-model profile, widely adopted in X-ray analyses.
\item Using the aforementioned profile, we have fitted deep \emph{Chandra} observations of
the X-ray surface brightness of MACS~0416.
We have found a best-fitting model consisting of two diffuse components with similar
values of $R_C \approx 180$~kpc for the core radius.
An additional, more compact, dPIE distribution is required to match the peaked
emission in the central regions of the northeast clump.
Our findings are in agreement with those of a previous work by \citet{Ogrean2015}
and with the results of a radial deprojection technique.
\item Once the intracluster gas is included as a separate mass component, the values
of the parameters inferred from the new lensing analysis are similar to those
obtained in the reference model (in all cases within the $3\sigma$
confidence levels, given the current model and data uncertainties).
Moreover, the total mass does not change between the two models and only the
diffuse halo contribution is reduced, by approximately $10\%$, while the mass
of the member galaxies remains the same.
\item Taking advantage of our new model, we have reconstructed the spatial distribution of
the total and intracluster gas surface mass density of MACS~0416, showing some
spatial differences between the collisional and noncollisional matter.
This provides a more self-consistent measurement of these two, intrinsically different,
mass components.
The measured offset, of about $6\arcsec$ in the southwest region, is consistent
with MACS~0416 being in the initial phase of pre-merging, as discussed in
\citet{Balestra2016}, with the gas components mildly trailing behind the
noncollisional components (stars and DM).
Furthermore, our method provides a possibly unbiased measurement of the center of
the dark matter distribution, a quantity that can be used to measure the cross
section of SIDM.
\item We have found that in MACS~0416 the projected fraction of total mass in
diffuse halos, composed mainly of dark matter, is almost constant in the region
from $\approx 70$ kpc out to more than $350$ kpc from the northern BCG.
This demonstrates the importance of modeling separately and
disentangling the hot gas component to measure more accurately the dark matter
distribution in galaxy clusters.
\item Finally, we have provided both the 2D map and the 1D cumulative profile of
the projected gas over total mass fraction.
From our model of the X-ray surface brightness, we have estimated that the projected
gas mass within an aperture of $350$ kpc is $M_\text{gas}(R<350\text{kpc}) = 3.3 \times 10^{13} $M$_\odot$
(with a few percent statistical errors), which gives a projected gas fraction of
approximately $10\%$.
\end{itemize}

The framework we have
presented combines X-ray and lensing observables in a more consistent way than
a posteriori analyses: this is a step forward in a broader effort to paint a
multiwavelength picture of clusters of galaxies, complementary to the other
joint techniques.
In MACS 0416, a simpler analysis, where the hot gas is subtracted from the diffuse
halo, \replaced{would have resulted}{results} in a similar cumulative mass profile for the
dark matter component.
Despite that, we have shown that our improved mass model can determine more
accurately the values of the parameters adopted to describe the inner
dark matter distribution of a cluster, thus providing more suitable
results to test different structure formation scenarios and the collision-less
nature of dark matter.

\acknowledgments
M.B. and C.G. acknowledge support by the VILLUM FONDEN Young Investigator Programme
through grant no. 10123.
S.E. acknowledges the financial support from contracts ASI-INAF I/009/10/0,
NARO15 ASI-INAF I/037/12/0 and ASI 2015-046-R.0.
G.B.C., P.R., A.M., M.A., and M.L. acknowledge financial support from PRIN-INAF
2014 1.05.01.94.02.
Corner plots were created using the \emph{corner.py} module \citep{Foreman2016}.

\appendix

\section{Posterior distributions}
\label{sec:posteriors}
\begin{figure*}
	\centering
	\includegraphics{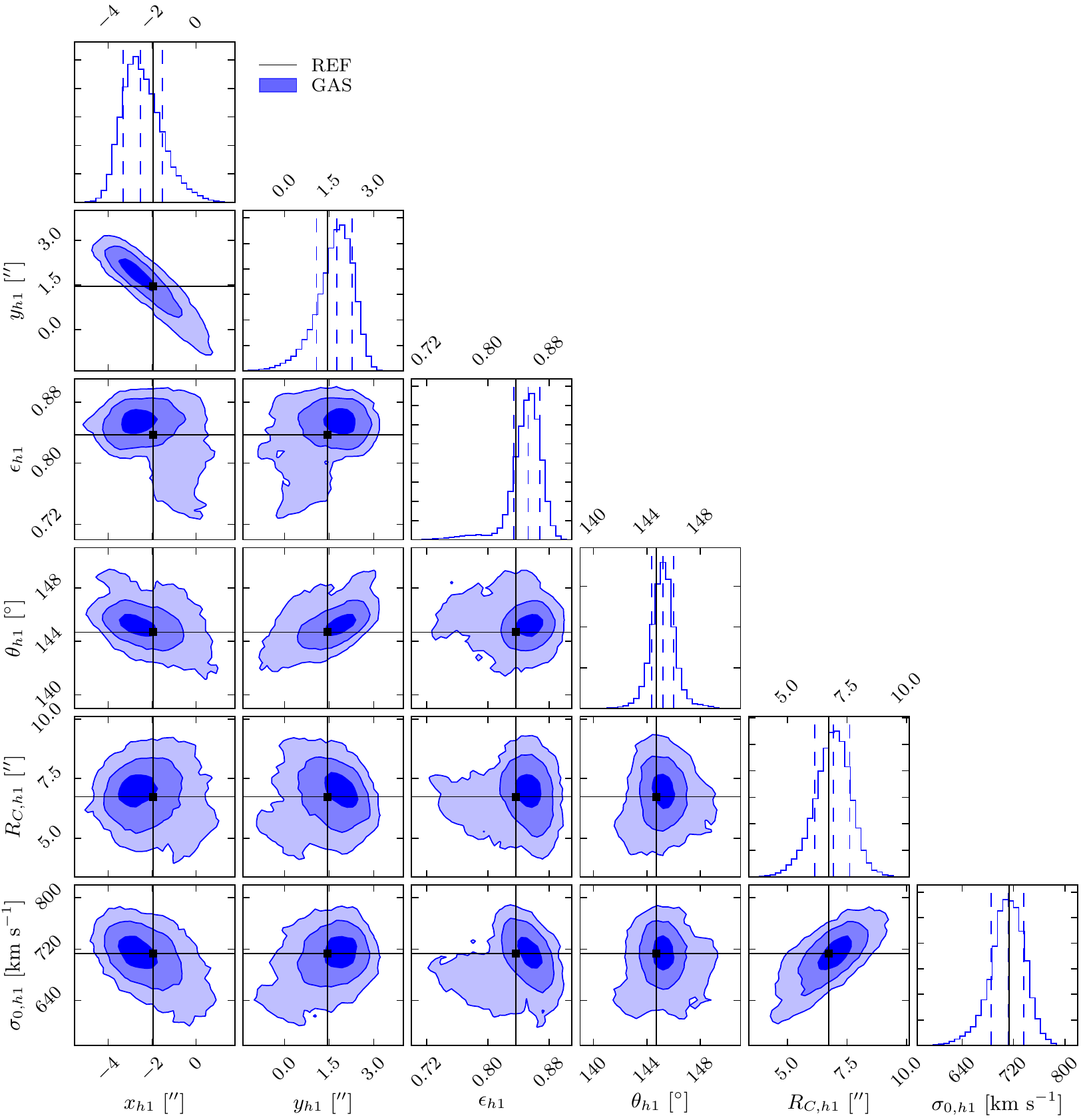}
	\caption{Posterior distributions of the values of the North-East main halo parameters
	derived from the strong lensing model with the hot gas included separately (i.e., GAS model).
	Blue contours contain $39.3\%$,  $86.5\%$ and $98.9\%$ of the samples,
	while blue dashed lines in the 1D histograms show the $16^\text{th}$, $50^\text{th}$
	and $84^\text{th}$ percentiles. Black solid lines mark the median values of the
	reference model (i.e., REF model).}
	\label{fig:corner_gas_comp_O1}
\end{figure*}

\begin{figure*}
	\centering
	\includegraphics{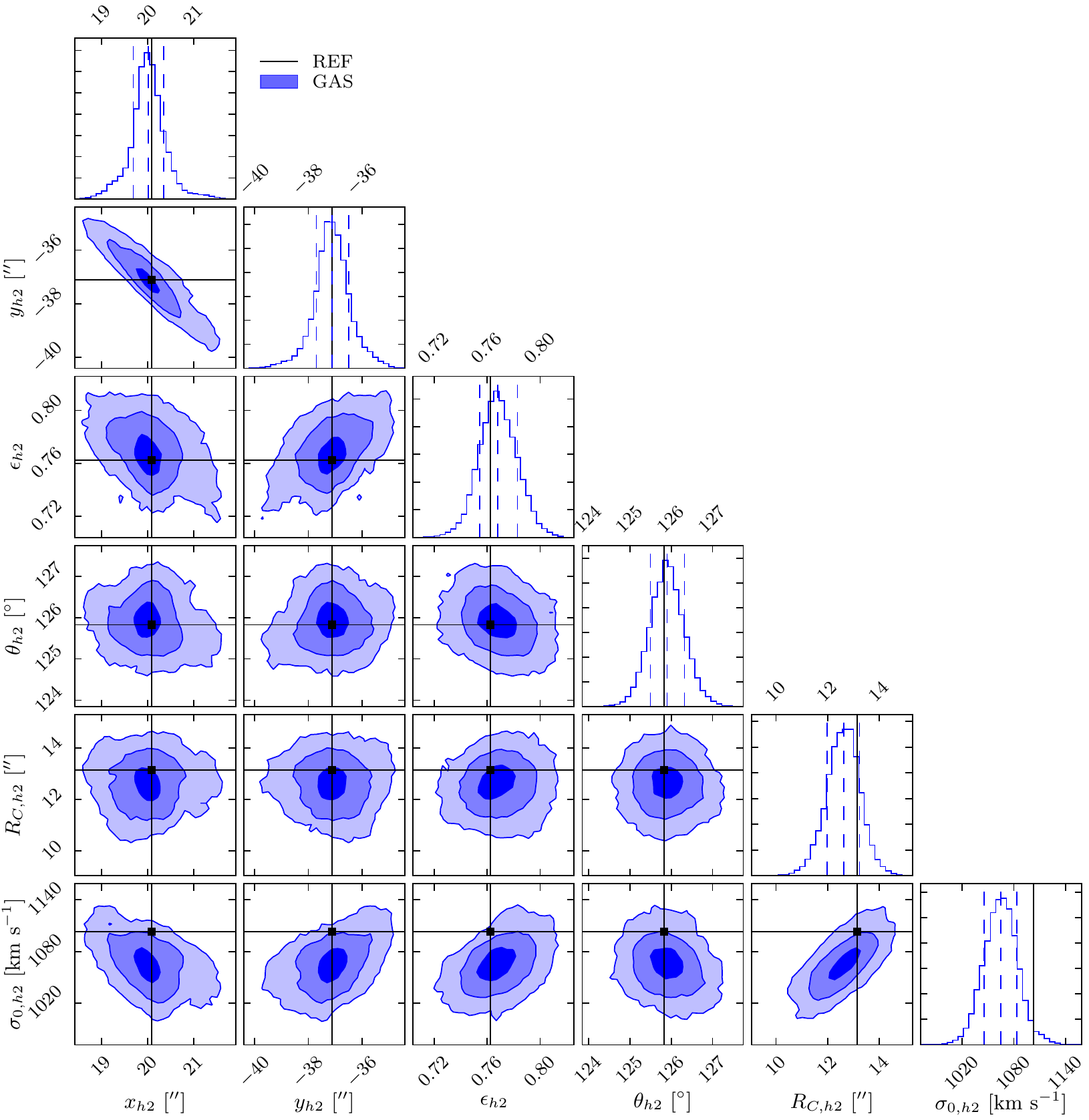}
	\caption{Posterior distributions of the values of the South-West main halo parameters
	derived from the strong lensing model with the hot gas included separately (i.e., GAS model).
	Blue contours contain $39.3\%$,  $86.5\%$ and $98.9\%$ of the samples,
	while blue dashed lines in the 1D histograms show the $16^\text{th}$, $50^\text{th}$
	and $84^\text{th}$ percentiles. Black solid lines mark the median values of the
	reference model (i.e., REF model).}
	\label{fig:corner_gas_comp_O2}
\end{figure*}

\begin{figure*}
	\centering
	\includegraphics{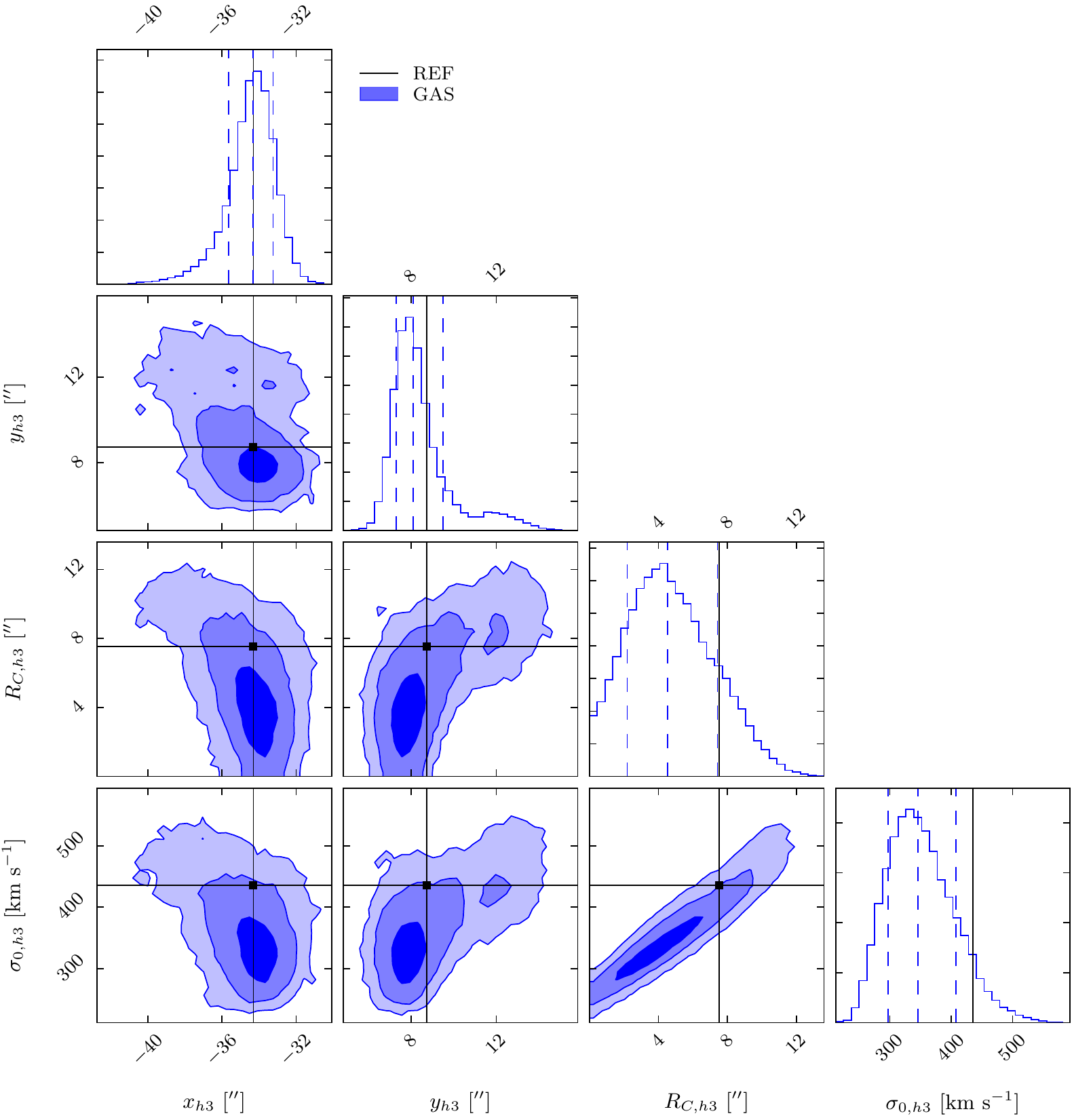}
	\caption{Posterior distributions of the values of the third main halo parameters
	derived from the strong lensing model with the hot gas included separately (i.e., GAS model).
	Blue contours contain $39.3\%$,  $86.5\%$ and $98.9\%$ of the samples,
	while blue dashed lines in the 1D histograms show the $16^\text{th}$, $50^\text{th}$
	and $84^\text{th}$ percentiles. Black solid lines mark the median values of the
	reference model (i.e., REF model).}
	\label{fig:corner_gas_comp_O3}
\end{figure*}

Hereafter (Figures \ref{fig:corner_gas_comp_O1}-\ref{fig:corner_gas_comp_O3}) we
show the posterior distributions of the parameter values of the main halos of MACS~0416
derived from the strong-lensing model with the hot gas included separately (i.e., GAS model).
As described above, blue contours contains the  $39.3\%$,  $86.5\%$ and $98.9\%$ of the
samples, which correspond to  the $1\sigma$, $2\sigma$ and $3\sigma$ of a 2D Gaussian
distribution, while the black solid lines are the median values from the
reference model (i.e., REF model).
In the 1D histograms, vertical blue dashed lines coincide with $16^\text{th}$, $50^\text{th}$,
and $84^\text{th}$ percentiles.


\clearpage

\section{Analytical solution for two-component spherical and ellipsoidal dPIE models}
\label{sec:d2PIE}
We can use the analytical solution presented in Equation (\ref{eq:dNPIE}) to
obtain explicitly the X-ray surface brightness of a system consisting of two
dPIE components:
\begin{equation}
\begin{split}
&S_X(x,y) \propto \int_{-\infty}^{+\infty}\!\!\! \rho^2 \text{d}z =\\
& = \frac{ \pi \rho_{01}^2 R_{C1}^4 R_{T1}^4 \left(  R_{A1}^2 + 3 R_{A1} R_{B1} + R_{B1}^2 \right)}
{ 2 R_{A1}^3 R_{B1}^3  \left(R_{A1} + R_{B1}\right)^3 } 
+ \frac{ \pi \rho_{02}^2 R_{C2}^4 R_{T2}^4 \left(  R_{A2}^2 + 3 R_{A2} R_{B2} + R_{B2}^2 \right)}
{ 2 R_{A2}^3 R_{B2}^3  \left(R_{A2} + R_{B2}\right)^3 } 
+ 2 \pi \left( \rho_{01} R_{C1}^2 R_{T1}^2 \right) \left( \rho_{02} R_{C2}^2 R_{T2}^2 \right)   \times \\
&\times \frac{ R_{A1}^2 (R_{B1}+R_{A2}+R_{B2}) + R_{A1} (R_{B1}+R_{A2}+R_{B2})^2
            + (R_{B1}+R_{A2}) (R_{B1}+R_{B2}) (R_{A2}+R_{B2}) }
            {R_{A1} R_{B1} R_{A2} R_{B2} (R_{A1}+R_{B1}) (R_{A1}+R_{A2}) (R_{A1}+R_{B2})
             (R_{B1}+R_{A2}) (R_{B1}+R_{B2}) (R_{A2}+R_{B2})}.
\end{split}
\end{equation}

Moreover, relaxing the spherical assumption and choosing an ellipsoidal mass
density distribution with two axes laying on the plane of the sky, the core and
truncation radii can always be rescaled as follows:
\begin{equation}
\begin{split}
\rho(x,y,z) = & \frac{\rho_0}{ \left( R_{C}^2 + \frac{x^2}{a^2} + \frac{y^2}{b^2} + \frac{z^2}{c^2} \right) 
                 \left( R_{T}^2 + \frac{x^2}{a^2} + \frac{y^2}{b^2} + \frac{z^2}{c^2} \right)} = \\
= & \frac{\rho_0 c^4}{ \left( R_{C}^2 c^2 + \frac{x^2 c^2}{a^2} + \frac{y^2 c^2}{b^2} + z^2 \right)
                 \left( R_{T}^2 c^2 + \frac{x^2 c^2}{a^2} + \frac{y^2 c^2}{b^2} + z^2 \right) } = \\
= & \frac{\rho_0'}{ \left( R_{A}'^2 + z^2 \right) \left( R_{B}'^2 + z^2 \right) }.\\
\end{split}
\end{equation}
From the substitutions in the last step and a comparison with Equations
(\ref{eq:dpie_3Ddens}), (\ref{eq:dpie_radii}) and (\ref{eq:I_i}), it is clear
that the solution for an ellipsoidal dPIE will be the same as that for a
spherical one, once the radii and densities are properly rescaled.

\bibliography{library}


\end{document}